# 4 Principles Fundamental to Design Practice for human centred systems

## Introduction

*" Designers of older technologies such as Postal Services, Telephones, and Television have reached the goal of universal usability, but computing technology is still too difficult to use for many people. One Survey of 6000 Computer users found an average of 5.1 hours per week wasted trying to use computers. More time is wasted in front of computers than on highways."*

Ben Schneidermann[1]

Poor design criteria are responsible for wasting computer users time and are a hindrance to effective interaction with human centred systems. It is important that before any design of an interface is attempted an in-depth analysis of task and user needs must be undertaken. Questions need to be asked such as "What takes place during interaction between the user and the interface?" "Does the interaction produce the desired result?" "Can the interface be easily changed to accommodate evaluation feedback?"[2]

**Analysis of the literature**

There is at present no complete taxonomy of design principles to aid us in deciding on the principles fundamental to design.

The first step therefore was to analyse and codify the design principles found in three of the main texts used, namely Dix et al, Schneiderman and Preece. This analysis is summarised in figure 1 and demonstrates that there is as yet no firm agreement as to the relative importance of design principles.

It is clear that Dix et al have produced the closest approach to a full taxonomy, but they do not claim exhaustiveness, which is an essential element to a full classification scheme. This is yet to be attempted.

**Choosing 4 principles**

The task is to chose and justify four fundamental design principles. This was done through a combination of processes involving discussion with colleagues, workers in the field, usability statistics obtained by a tailored questionnaire from a random selection of users asked to evaluate a specifically designed web site, and by research and analysis of the literature available.

The Four design principles chosen are

| | |
|---|---|
| **1 Learnability/Familiarity** | **2 Ergonomics/Human Factors** |
| **3 Consistency/Standards** | **4 Feedback/Robustness** |

| Dix et al | | | Schneidermann | Preece |
|---|---|---|---|---|
| Learnability | Predictability | Operation Visibility | | |
| | Synthesizability | Effect of past Ops on current State | | Ensure ease of understanding |
| | Familiarity | Guessability Typewriter Metaphor | Principle 8 Reduce short term memory load | Use appropriate metaphors |
| | | | | Reduce cognitive load |
| | Generalizability | Cut&Paste in Multiple Windows Ops | | |
| | Consistency | Likeness in behaviour | Principle 1 Strive for consistence | Maintain consistency and clarity |
| Flexibility | Dalogue initiative | User pre-emptive | Principle 4 Design Dialogues for closure | Allow input flexibility |
| | Multi-threading | Simultaneous communication | | |
| | Task migratablity | Control transfer of task execution | | Design for user growth |
| | Substitutivity | Representation multiplicity | Principle 2 Enable short cuts | Provide shortcuts |
| | Customizabiltiy | Adaptivity | | Adapt to different user levels and styles |
| Robustness | Observability | Browsability | | |
| | | Use of Defaults | | |
| | | Reachability | Principle 7 Support internal locus of control | |
| | | Persistence | | |
| | Recoverablility | Forward error recovery | Principle 5 Error Prevention | |
| | | Backward Error recovery | Principle 6 Easy reversal of actions | |
| | | System Recoverablity | | Engineer for errors |
| | | User Recoverability | | Provide a 'RESET" command |
| | | Commensurate effort | | |
| | Responsiveness | Response time | Principle 3 Offer informative feedback | Give appropriate quantity of response |
| | | Stability | | |
| | Task conformance | Task completeness | | |
| | | Task Adequacy | | |

*Figure 1: Analysis and comparison of principles determined from Dix, Schneidermann and Preece*

The appropriateness of the four principles can be illustrated through the Abowd and Beale model of interaction.

The four principles cover the full range of System/User/Input/Output Interaction as shown in the table below:

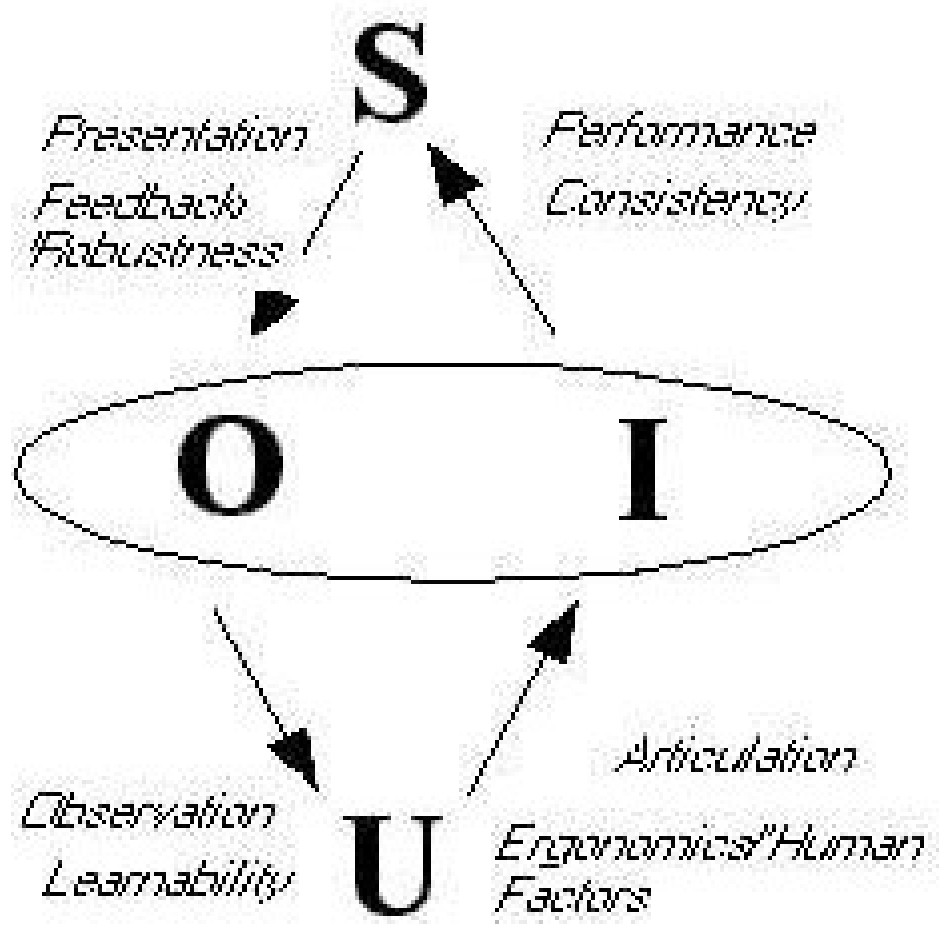

*Figure 2*

| | SYSTEM | USER |
|---|---|---|
| INPUT | **3 Consistency** | **2 Ergonomics/Human Factors** |
| OUTPUT | **4 Feedback/Robustness** | **1 Learnability/Familiarity** |

Expanding the Abowd & Beale diagram we can apply it directly to our own four principles, which are adapted, amplified and modified from Hewett et al[3] into to our own scheme and shown in figure 3 below.

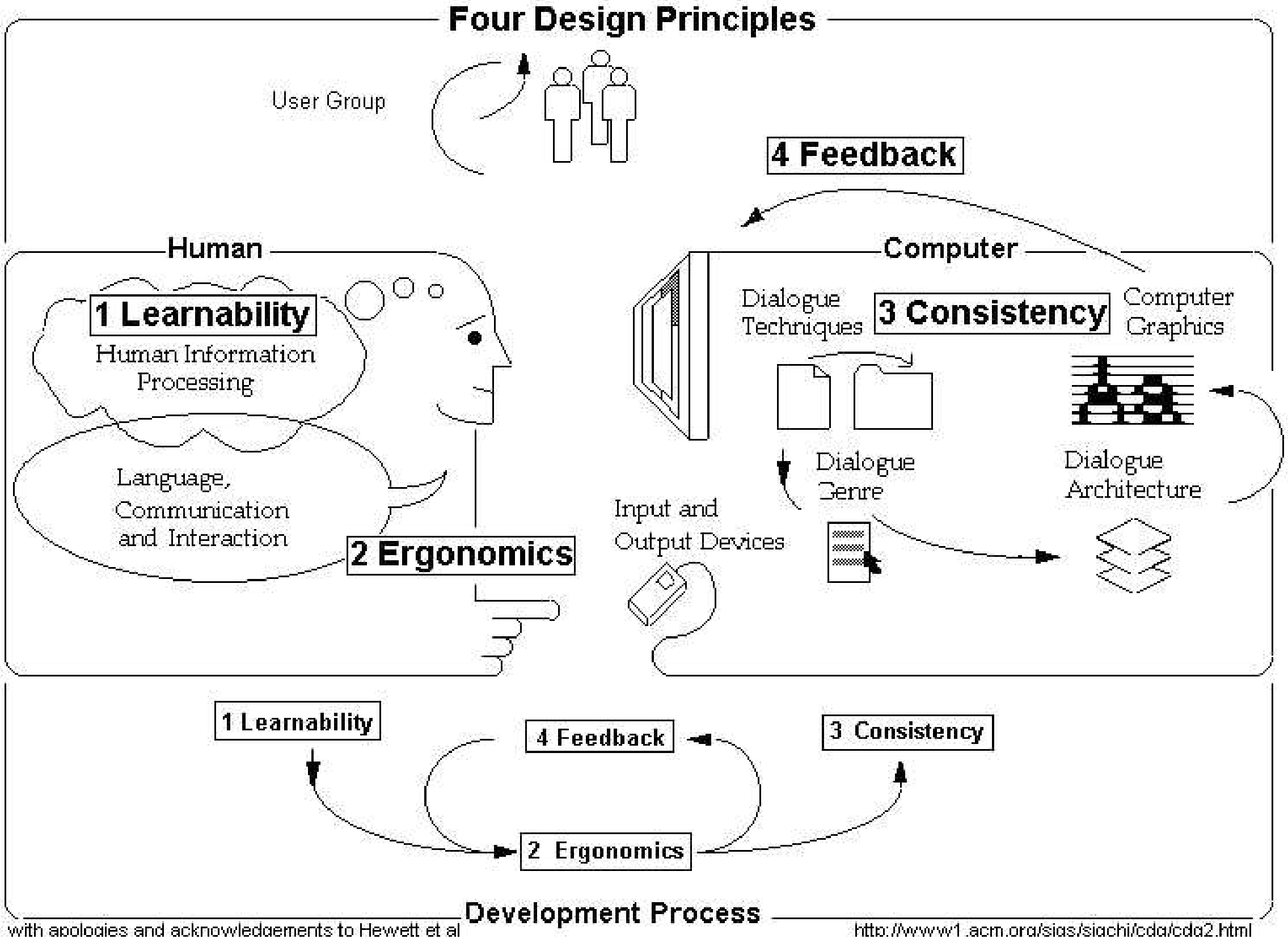

*Figure 3: The four chosen principles*

The four principles span the entire locus of interaction from intellectual conception within the human brain (learnability) through physical and immediate interaction with the interface (ergonomics) into dialogue architecture within the system (consistency) and out and back to the user through visual and other stimulus (feedback/robustness)

These four principles encompass a full 4-stage feedback loop covering the range of human centred interaction. **1)** learnability informs **2)** ergonomics, which informs **3)** consistency, which provides **4)** feedback. An inner feedback circle exists where stages 2) and 4) inform each other to provide an evolutionary robustness. For example a fast moving mouse pointer on the screen can be adjusted to provide easier control for the hand, which provides clearer feedback, and better control and so on demonstrating an almost Darwinian evolution of user features progressing towards greater usability.

These four principles work in harmony over the whole range of human system interaction. Each of the principles are now addressed in turn

## Principle 1: Learnability/Familiarity

*"What users want is convenience and results"*
J.Raskin[4]

The principle hindrance to effective interface interaction is lack of familiarity based upon knowledge of the system. The longer the time taken to acquire that familiarity the more difficult it is for the user to interact with the interface effectively. This learning time can be reduced by making use of the knowledge that the user already has.

When the user performs an unfamiliar task there is an inherent internal barrier to getting the task completed. The learning process has to become involved in order to accomplish the user-centred goal. Any system, which reduces or obviates this learning process will enhance task performance by reducing mental workload.

An example of this is the payphone coin tray shown below which illustrates making a function visible and instantly intuitive by simply placing the coins in the slide on the top of the machine. Coins fall one by one only when needed. This leaves the user in no doubt how much money is left and provides a simple mechanism for inserting coins without continually fumbling for change. Figures 4 and 5.

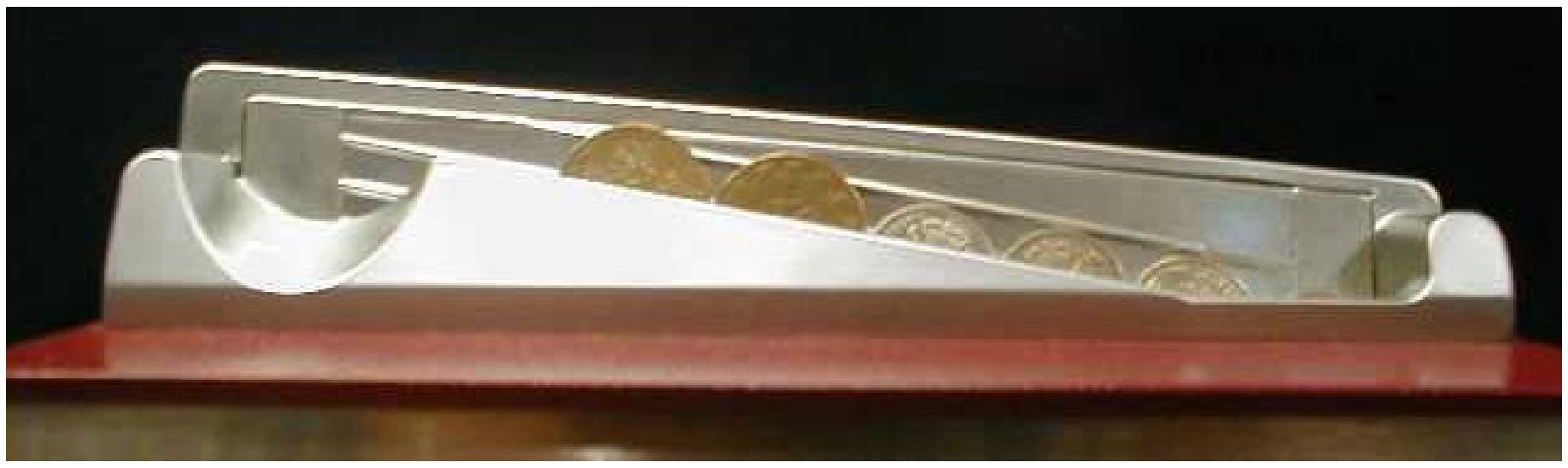

*Figure 4 www.eit.ihk-edu.dk/subjects/*

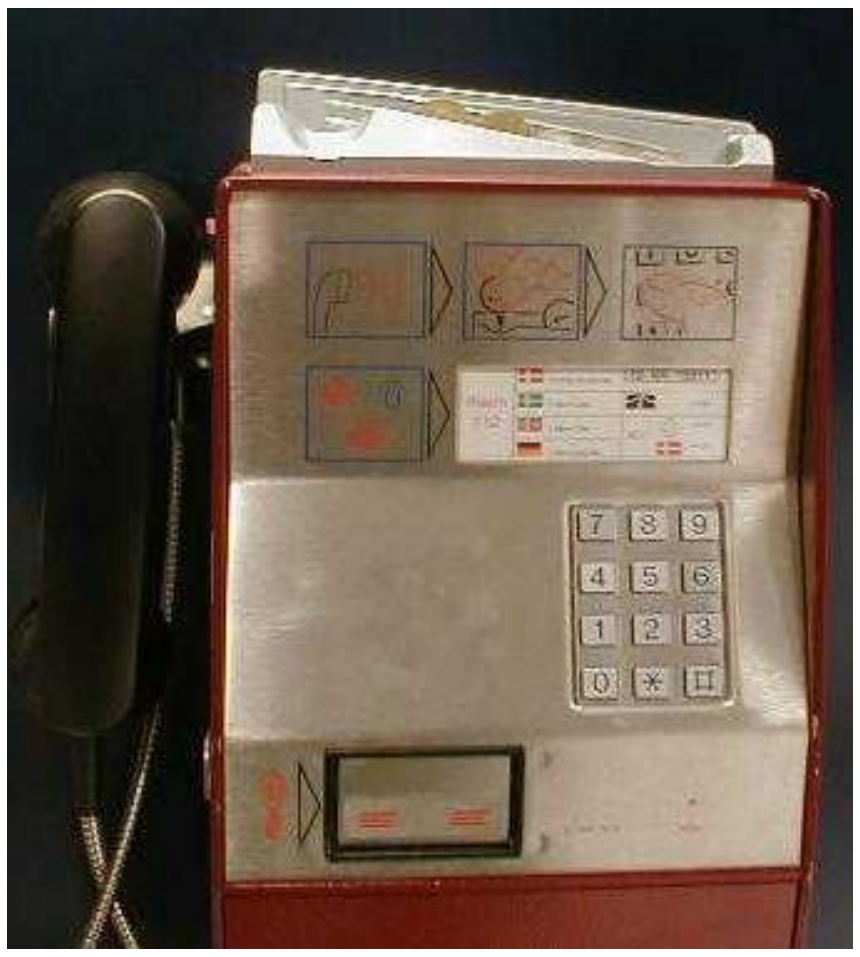

*Figure 5www.eit.ihk-edu.dk/subjects/*

If it is possible to minimize or eliminate altogether the learning process, the results are obtained quickly and conveniently, and according to Raskin this is the core of what users want.[5]

One common way of applying learnability is to use a paradigm known to the user. Common themes such as office or desktop can utilize the users prior knowledge. Similarly screen objects can suggest by their colour, shape and other attributes what can be done with them. This is related to the idea of **affordance**,[6] where the intrinsic property of an object or group of objects suggests how they should be manipulated. Affordance refers to ... "*the actionable properties between the world and an actor*" J.J. Gibson 1977/1979. To Gibson affordances are a relationship " *They are a part of nature; they do not have to be visible, known or desirable.... some are dangerous. I suspect that none of us know all the affordances of even everyday objects*."[7] Donald Norman.

Common Paradigms that have been used for interface modelling include the Office Desktop where the Filing cabinet, Calculator, Briefcase and the Recycle Bin have an immediate recognition and usage. This desktop metaphor, which has been with us for over 20 years, and is the most widely used and recognised is overdue for revision. There have been many suggested extensions, the most obvious of which is the 3D extension, which enables the user to locate objects in a 3D world[8]. Alternatively Hutchings and Stasko[9] have developed a new set of intuitive operations for expanding windows utilisation.

Making the system easy to learn provides the user with predictive capabilities and cognitive understanding of the methods of accomplishing the task. If you can predict what comes next you do not have to learn it.

Our choice of learnabiltiy as one of the four fundamental principles is justified on three grounds:

**Justification 1** springs from an analysis of the literature.

Dix et al (figure 1) use the term 'Learnability" as one of its top -level categories. Schneidermann's principle 8 and Preece's (reducing cognitive load) and (ensuring ease of understanding) further support its importance.

Furthermore learnability through common paradigms is considered valuable. *"Tremendous commercial successes in computing have arisen directly from a judicious choice of metaphor..... Very few will debate the value of a good metaphor for increasing the initial familiarity between the user and the computer application"*[10]

**Justification 2** is generated from within the group's discussions on 28th Oct.2002

> **Author**: Samra, Jasmine <jms601@soton.ac.uk>
> **Subject**: Assuming on the website"
> "We fouind that whilst surfing through websites, ASSUMING played a large part in how the user thought they understood how to use the website correctly. [11]

The notion of assumption is closely related to that of expectation and the group idea is that correct usage of the interface flows from correct expectations. If the assumed expectations are incorrect the user is led astray. It is therefore essential to build in supports for correct expectation to reduce the learning curve.

The group further felt that familiarity was a key concept for the usability of a Website, and without it new users could easily be entirely lost. Lack of familiarity is equated with a lack of user friendliness.

> **Author:** Keshani, Shilain <ssk85@hotmail.com>
> **Subject:** New User???
> As others in my group were familiar with the website i was navigating, and i was not, it made them realise that the website was not as user friendly as they initially thought. Things that seemed quite obvious to them didn't seem simple to a new user. [12]

This is a common problem for all interface designers – anticipating what the user does *not* know and not assuming that all users will have the level of knowledge required to be utilized successfully. Of course, as it stands such comments do not go deep enough. The question of what user friendliness really is, is left unanswered and needs more definition.

**Justification 3** is based on our own user survey. See Appendix 1

A questionnaire was devised to ascertain the usability statistics and user views of the design principles engineered into the Website. Questions 1 and 3 dealt with the issue of learnabilty and 80% of users agreed or strongly agreed that the Website navigation was intuitive and familiar. While the survey users placed 'easy to learn" as the 3rd most important principle in navigating an unfamiliar Website.

**Application to Recipe Website**
We have deliberately chosen commonly used devices such as the Calendar to depict seven daily choices, and a menu to relate these choices to the Lunchtime recipe, which is the basis of the Web site. This is engineered into the Lunchtime Recipe Website with a daily toolbar, which provides immediate navigation to the recipe of the day. If the user clicks on 'Tuesday" he can expect to get the recipe for Tuesday. This is designed to be obvious to the user to the point of automatically triggering the cognitive consciousness of the human mind to spontaneous intuition.

Furthermore investigation of a variety of common recipe books indicated a standard layout adopted by most authors, which involved three components.

**1. A picture of the dish**

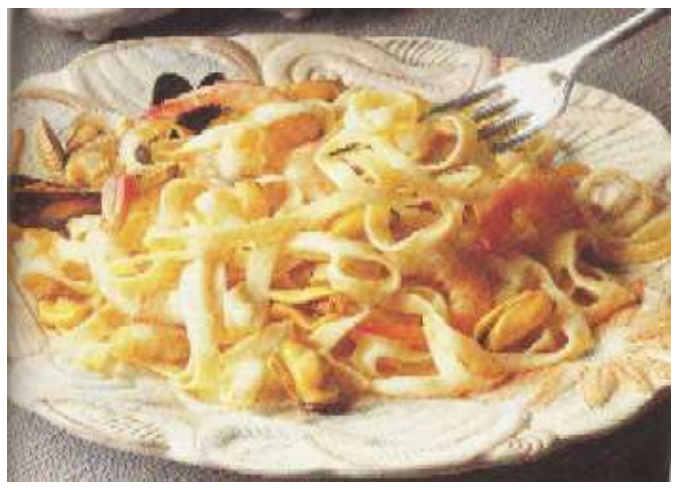

**2. A list of ingredients**

| INGREDIENTS |
| --- |
| **Seafood Tagliatelli**<br>**Serves 2 \| Cooking time 20 minutes**<br>**Ingredients**<br>8 oz Tagliatelli pasta<br>4 oz drained tinned tuna<br>12 cooked prawns<br>1 tin of drained mussels<br>4 oz cambozola cheese<br>2 oz butter<br>5 fl oz single cream<br>4 cloves crushed garlic<br>5tbsp fish stock<br>3 tbsp white wine<br>1 tbsp olive oil<br>salt, pepper and paprika. |

**3. Description of the method**

| METHOD |
| --- |
| **Method**<br><br>**1**. Boil the tagliatelli in plenty of boiling salted water with the olive oil added and cook for 10 minutes or until just al dente. Drain well.<br>**2.** Melt the butter in a saucepan large enough to hold all the ingredients and add the drained pasta and the garlic. Stir thoroughly, add the white wine, and cook for 1 minute.<br>**3.** Add the cheese and stir well until the cheese has melted then add the cream<br>**4.** Add the mussels, prawns, tuna and fish stock and stir until all ingredients have been heated through well.<br>**5**.Season to taste and put into a warmed serving dish, sprinkle with the paprika.<br><br>Serve with whole leaf spinach or a mixed salad<br><br>**Variation** This dish can be made with flaked smoked haddock or salmon instead of the tuna |

By adopting the cookery book paradigm the users' prior knowledge is again utilized.

## Principle 2: Ergonomics/Human Factors

*"ERGONOMICS: Anthropometric and physiological characteristics of people and their relationship to workspace and environmental parameters"*

ACM SIGCHI Curricula for Human Computer Interaction[13]

According to Dix et al ergonomics is not strictly HCI. This is disputed however by *ACM SIGCHI Curricula for Human Computer Interaction* identifies ergonomics as part of HCI. (See Appendix 2) It is therefore argued that Human Computer Interaction cannot take place without taking ergonomics into the equation. Although the literal meaning of ergonomics is "energy efficient" Ergonomic design involves understanding the users and what they need to do with the system and where they will use it. When the user is seen as an unreliable, unpredictable component of the system, the importance of the user's humane characteristics is negated and undervalued, resulting in systems that are difficult to use, awkward and inefficient.

Failure to prioritise ergonomics/human factors as a fundamental design principle has led to failures we have all experienced.

- The keypads (some Mobile Phones) designed for miniature fingers,
- The displays on LCD and similar display screen equipment (PDA's) which become illegible in daylight,
- Illogical operating sequences (some VCR's) which have not been tested on real people.
- Cluttered Lottery ticket design buried important wrong date information in the middle of the ticket, leading to loss of £1.6m for the customer.[14]

Olinka Koster[15] reports in a recent article on the findings of Chiba University in Japan that sitting in front of a computer screen for extended periods can dramatically increase the risk of depression and insomnia. Psychology professor Cary Cooper from the University of Manchester Institute of Science and Technology is quoted as saying that concern was growing over mental health problems caused by working with computers.

Individual users possess common capabilities such as reading, listening, touch, and motor skills, in addition to average height, size and weight, the physical diversity, psychological abilities and disabilities of users have to be analysed and evaluated when designing interfaces e.g. switches that need to be operated in sequence need to be placed together.

Car manufacturers have had to pay particular attention to the issue of ergonomics in order to make their products simple to operate intuitive to use and above all safe to drive. However even the design of an item such as the car radio can have a big impact on the whole driving experience.

David Travis in his article" Car Radios: It' s time to get Angry" www.Systems-concepts.com[16] tells how it took him two days and a friend's translation to work out how to tune the radio to the next station. The consequences of this kind of frustration in a driving environment can be dangerous to the extreme. Whatever car Mr Travis hired he found he could always familiarise himself with the driving controls in under a minute, but it was quite different with the radio. "*I have still to find a single radio that I can use with "one trial learning". Imagine if you had to spend your first thirty minutes sitting in your new rental car working out how to drive it away. There would be uproar!"*

On the other hand the Sony 2010 multiband radio achieves ease of use by providing 32 dedicated buttons for pre-sets. These pre-set buttons are not specific to one particular band, which means that a single press of a button will take you directly to the radio station of your choice. Jef Raskin[17] says of this radio *"It is rare when a product combines both superior electronic technology and interface technology. Perhaps this explains why the Sony 2010 has been in continuous production for over a decade in an Industry where most models fade in months."*

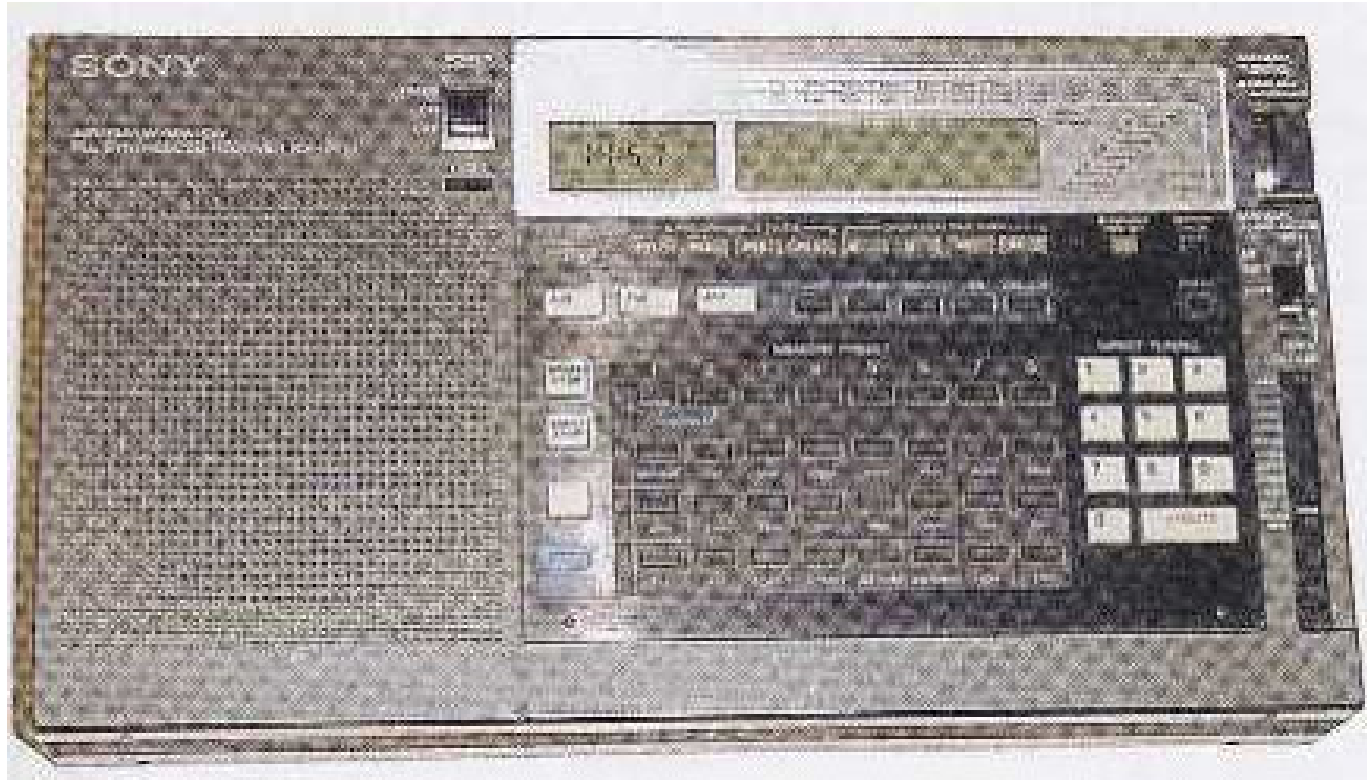

*Figure 6 Jef Raskin, The Humane Interface,P135*

This Sony radio illustrates one of the most useful ergonomic applications that can be applied to the web page environment, namely **one-click navigation**. If it is possible to get to any part of the web site with a single click this reduces the amount of work to the minimum. In the Recipe web site we have implemented the **one-click navigation principle** as a design feature.

The amount of work done in moving the mouse around a particular application is dependant upon the average time it takes a user to succeed in getting the cursor to the button or icon etc. This can be measured quantitatively using Fitts' Law which states that in one dimension:

$$\text{time (in m sec)} = a + b \log_2 (D/S + 1)$$

where **S** is the size of the target, D is the distance of the target from the starting point and **a** and **b** are constants determined by experiment. The smaller the value the better the design and the more ergonomic the interface.

Complete navigational freedom is one of the goals of ergonomic design. All too often the user is controlled by the system instead of controlling the system. The following form filling situation is unfortunately all too common. The user is asked to enter his name and other personal details such as account and fax numbers which he does not want to give, and enters '0' instead, only to get replies such as illegal entry try again etc., often entering into a loop which he cannot escape from resulting in the user turning off the computer out of sheer frustration and losing data.

On the recipe web site the deliberate design goal has been to provide the user with full control. In particular the form can be filled in partly if the user so requires.

Our choice of ergonomics of one of the four fundamental principles is justified on three grounds

**Justification 1**. By the importance given to it within the research literature:

"*Properties of human learning and performance are directly applicable to the foundations of any Interface design.*"[18]

"*People cost a lot more money than machines, and while it might appear that increasing machine productivity must result in increasing human productivity, the opposite is often true.*"[19]

"*Inappropriate placement of controls and displays can lead to inefficiency and frustration... We can therefore see that appropriate layout is important in all applications.*"[20]

We conclude that ergonomics has a premier place in the efficient design of any Human centred system and we have to incorporate ergonomic and anthropometrical principles in the design of the recipe Website.

**Justification 2**. By Class room seminar and discussion.

The classroom where the seminars are conducted have served to provide a dynamic illustration of the importance of ergonomic principles.

The whole class has raised the issue of the use of a partially public classroom, which is unsuitable for learning due to noise pollution from a cognitive ergonomic point of view, and have thereby illustrated the importance of cognitive ergonomic issues to be considered in the design of any system for the delivery of information. This forms an eloquent underpinning of our contention that ergonomics should be one of our fundamental principles.

**Justification** 3. User Survey

Users were asked to place 8 design criteria in order of importance. Criterion 8 "least amount of work to achieve results" was the second most important principle from our survey. Furthermore 100% agreed or strongly agreed, 'One-click could get me anywhere I wanted" and a further 80% disagreed or strongly disagreed that this Website 'was hard to use".

**Website Application of Ergonomics**
It is clear that respondents valued features, which reduced their workload in accomplishing tasks. Features such as the print button and the **one-click navigation bar** (the user can move to any web page from any other web page on this site with a single click) performed tasks with fewer keystrokes and without having to enter a menu / dialogue box were included as part of the Website to meet this requirement.

Three buttons/links were constructed Print Button, Email Link and Submit Questionnaire. The Print Button enables the user to print out a recipe. Since we expect a number of our web pages to be printed out for use as working recipes we have employed the one click printer button on all recipe pages.

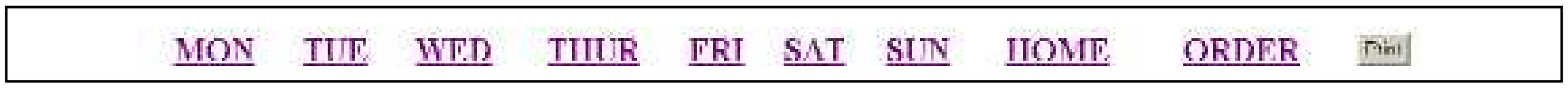

The print button is 2 cm wide and a distance of 14 cm from the centre of the screen. Applying Fitt's law, time (in m sec) = $a + b \log_2 (14/2 + 1) = a + 3b$, Setting the constants to default values of a=0 and b=1 provides a value of 3ms.

The email button enables the user to submit recipes. The Questionnaire button enables the user to provide feedback on the Website. This provides a one-click solution to all these tasks.

The WebPages were designed on a page per view basis, which enables the user to see a whole recipe without having to scroll, thus minimising the amount work required to see the recipe.

## Principle 3: Consistency/Standards

*"Consistency is a strong determinant of success of systems."*

Schneiderman[21]

Consistency and standards are central to usability. "Users should not have to wonder whether different words, situations or actions mean the same thing" Jakob Nielsen [22] Lack of consistency has even lead to dire consequences e.g. positional exchange of gun trigger and ejector control in Spitfire Aircraft, resulting in pilots ejecting rather than firing during dog fights in World War 2[23]

There are therefore different kinds of consistency on different levels. Consistency is required from cross platform consistency right down to individual user response.

The highest-level consistency within Human Centred Interfaces is consistency with user expectations (Level 1 – see figure 6). Consistency must be maintained across the range from the widest to the narrowest areas whether we are designing the interface for a mobile phone or for a suite of Office Applications.

Invisible structures (Level 2) refer to elements of the interface, which are hidden but provide utility to the user whilst small visible structures (Level 3) refer to buttons icons etc. Single application consistency (Level 4) requires consideration of splash screens, toolbars etc. Suites of interacting products (Level 5) need to share the same look, feel and operation across the whole range of units. Separate Hi-Fi units should have the same type of buttons in similar places. In-house consistency (Level 6) requires every product from the same manufacturer to share the same inherent design features, while platform consistency (Level 7) requires products from different manufacturers the meet the same standards.

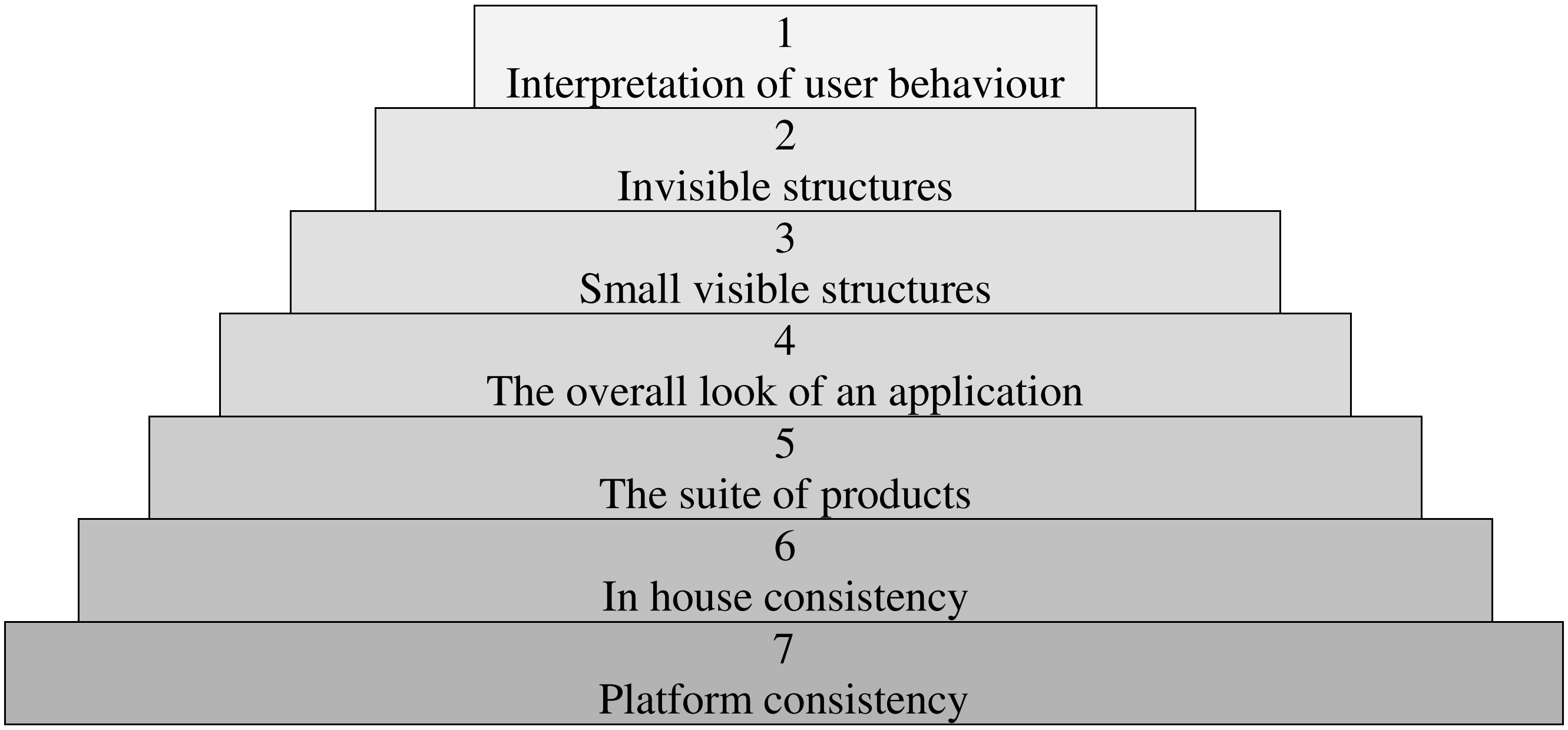

*Figure 7 Levels of System Consistency*

Global consistency depends upon standardisation. A volume control is always turned clockwise for every device. This universal standardisation means that no instructions are needed and no mistakes are made during operation.

For instance a car manufacturer knows that his next model has to place the clutch, brake and accelerator pedals in the same order as before to maintain safety standards.

The activating and terminating telephone buttons/icons on a, say, Nokia mobile phone have to be positionally consistent with previous models to maintain user expectations especially as upgrades are obtained frequently in the mobile communications area.

**Standards – the codification of consistency**
The design of a control panel or interface requires a check to see if there is a standard for colour, shape and placement.

Software standards provide consistency from one application to another. Dialogue boxes, for example, look the same in different windows applications. Minimisation /Maximisation buttons are all standardised in the Operating system rather than the application.

This consistency provides user advantage because a new program can be utilised immediately without having to learn how to save a file.[24] WebPages although less standardised often require a high degree of standardisation because of their diversity and multiplicity. For example hyperlinks are coloured blue and underlined as standard indications of an active link.

Different kinds of standards bring consistency in different areas

1. Secret factory standards - e.g. product compatibility
2. Proprieartry standards - e.g. Adobe portable document file
3. De facto standards - e.g. printer control languages
4. Official standards - e.g. HTTP Protocol and HTML language
5. Open Source - e.g. Unix and Linux

Of all these standards open source is the most useful to the user. Secret and proprietary standards make the user Company dependant. De facto and official standards are better but sometimes slow in being established and open source is an ongoing evolving standard open to all.

According to Cummaford and Long, *"There is a need for more formal HCI design knowledge, that is, whose conception is coherent, complete and fit for purpose, such that guarantees may be developed and ascribed"*[25]

Cummaford and Long's goal is the full implementation of a standard that can be used as a guarantee of a systems 'fitness for purpose' – a kind of universal HCI Kite-Mark that provides a stamp of usability on any system.

However such a standard is still a long way off. A full taxonomy of HCI principles has not yet been attempted. Only when it has been done and been accepted would Cummaford and Long's HCI guarantees be a viable prospect.

Our choice of consistency of one of the four fundamental principles is justified on three grounds

**Justification 1.** Course/Study Literature
According to Dix, consistency is one of the most mentioned principles on user interface design. "*However it has to be defined relative to something else*" – by which Dix means that consistency is transitive, it does not exist on its own, it is consistency with something else - *"preferably relating to the users' real world experience, and or consistent with past actions*"[26]

Schneiderman's view, quoted at the start of this section, is that *"Consistency is a strong determinant of success of systems."*[27]

Consistency has a foundational role in the efficient design of any Human centred system and we have incorporate consistency principles in the design of the recipe Website.

**Justification 2.** Research literature
ACM SIGCHI Curricula for Human Computer Interaction, Hewett et al have shown that the growth of discretionary computing and the mass personal computer market has resulted in the "the gradual evolution of a standardized interface architecture"[28] and along with these changes researchers and designers have developed *"specification techniques for user interfaces"* as a means of standardizing the design approach and providing consistency of operation across platforms. This is thought to become increasingly important, as the future will bring a wider spread use of computers by those who are outside the standard computing environment.

**Justification 3.** User Survey

Respondents placed criterion number 6, which deals with consistency, as the fourth most important principle in the design of a Website. Furthermore in the features comparison question a ratio of 9:1 respondent's favoured "consistent layout" to one "full of surprises".

In addition question 10 dealt with the consistency in design and 70% of users agreed or strongly agreed that the recipe Website met this criterion.

**Web site application of Consistency**

The principle of consistency has been applied to the recipe web site by ensuring the same "look and feel" to all of the seven rec ipe web pages. Each recipe page has the same layout, navigation bars, headings, graphics positions and print button placement.

The universal standard colour for web page links has always been blue. It was therefore felt that the web page links should maintain consistency with this standard and would strive for consistency of appearance and style.

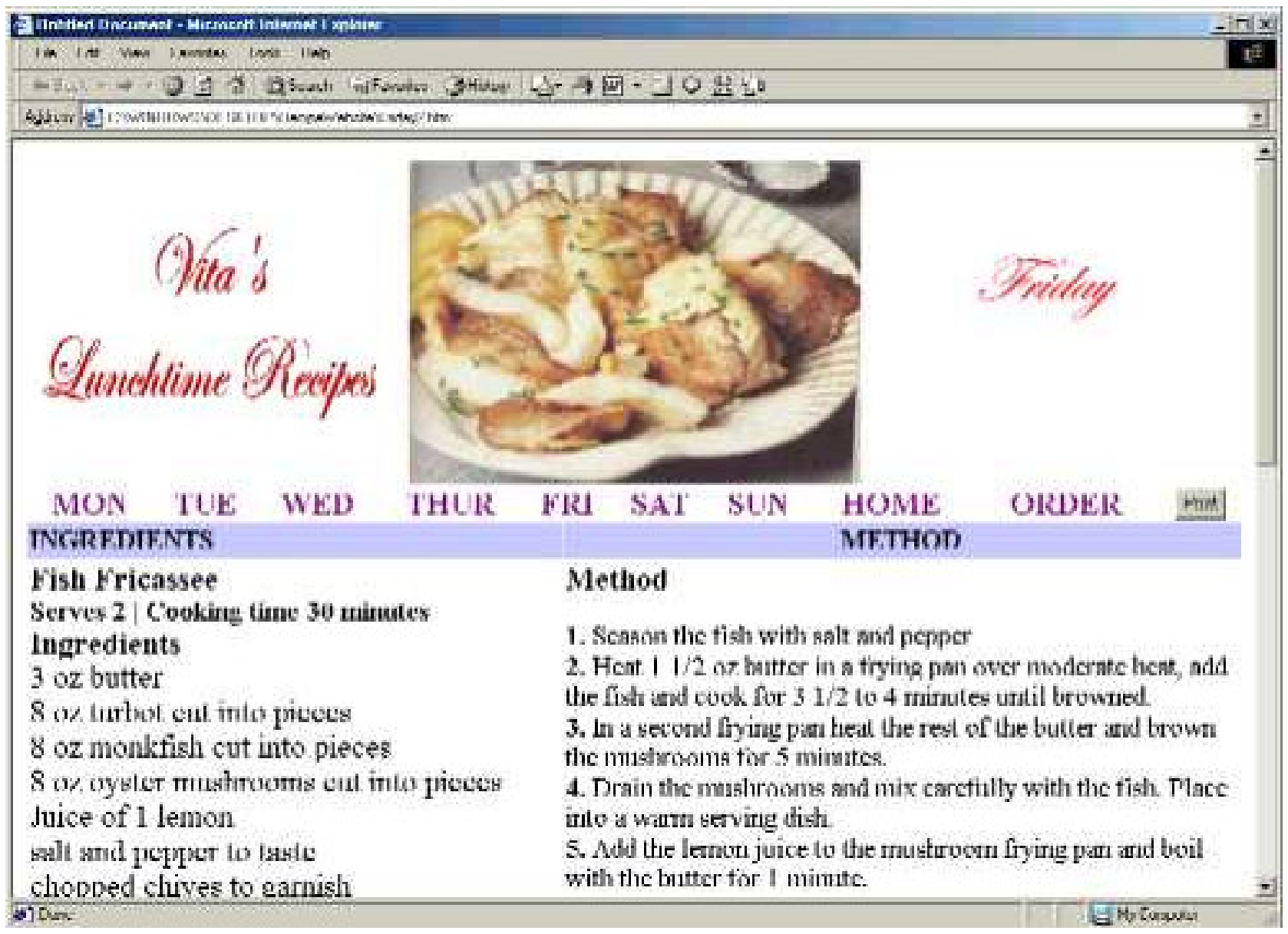

## Principle 4. Feedback / Robustness

Whenever a user operates a switch, presses a button, turns a dial, clicks a mouse or interacts in any way with a machine or with a system, there must be feedback that is unambiguous.

Feedback can come in the form of a sound, a light, text on a screen, dialogue boxes etc. However the ideal form of feedback is allowing users to see things happen.

Manual controls may provide visual feedback by the position of the switch (up or down), the pointer on a dial, a bank of flashing lights etc. Controllers such as those for operating DVD's on PC's, now consistently provide visual feedback on screen for the full range of operations from interactive participation, fine tuning to volume, colour and channel controls etc.

If a system gives no feedback at all then the user may assume that the command has not been received, accepted and acted upon. This may cause the user to press the button again, repeat the command or even think that the device is no longer working. Sometimes this may result in mistaken data input, endless looping, or a number of duplicated actions, which take further time and effort from which to recover.

The common method for informing the user that a process is undertaken is the use of the hourglass metaphor, which replaces the mouse pointer. However this provides only the minimum amount of feedback as it informs the user that a process is ongoing but not what that process is, neither what the process is working on nor how long it will take. Sometimes a progress bar is used for this purpose.

Feedback has become more sophisticated with enhanced intelligent user interfaces that **auto-complete** many of the inputs that humans would provide. However problems can arise when the computer tries to guess what the user wants but is not programmed sufficiently well. For instance many word processors provide an auto-complete feature where the program tries to guess the date or a salutation and which can often be inappropriate, forcing the user to take remedial action.

One of the characteristic uses of feedback is to alert the user to errors. Human beings can generally cope with simple errors quite easily. If a sentence is missing a word or even a comma it can usually be read easily enough. However computers are not so forgiving and the smallest error in syntax can produce an error response. A colon instead of a semicolon could produce incomplete data from a database or provide the wrong feedback to the user.

Making an interface error-responsive is a feedback issue. Even the simple misspelling of words in a word processor can be addressed by immediate feedback such as a red underline to provide the user with the possibility of immediate remedial action. The additional feature of substituting the correct spelling at the single click of a mouse button enhances the feature.

Habitual actions through task repetition can easily cause users to make mistakes when the locus of attention is shifted.[29] To allow users to undo those errors, even if they have already moved several steps ahead could improve error handling and recoverability.

The strength of a feedback should reflect the importance of the action. Sound and colour coding can be used here to indicate levels of information, warning, and danger. The introduction of a gradient warning sound depending on the grievousness of the error e.g. Fire alarm/ very serious; Ambulance siren/serious; Alarm clock/ wake up!! Early on in the error production, together with flashing lights and appropriate message boxes could enable to user to react before the damage is too great.

However when designing error prevention mechanisms for human centred systems according to the level of absorption or disturbance, error messages can get ignored if the computer behaves unexpectedly because stress level increases and the user is less likely to see help messages or react appropriately to warning sounds[30]

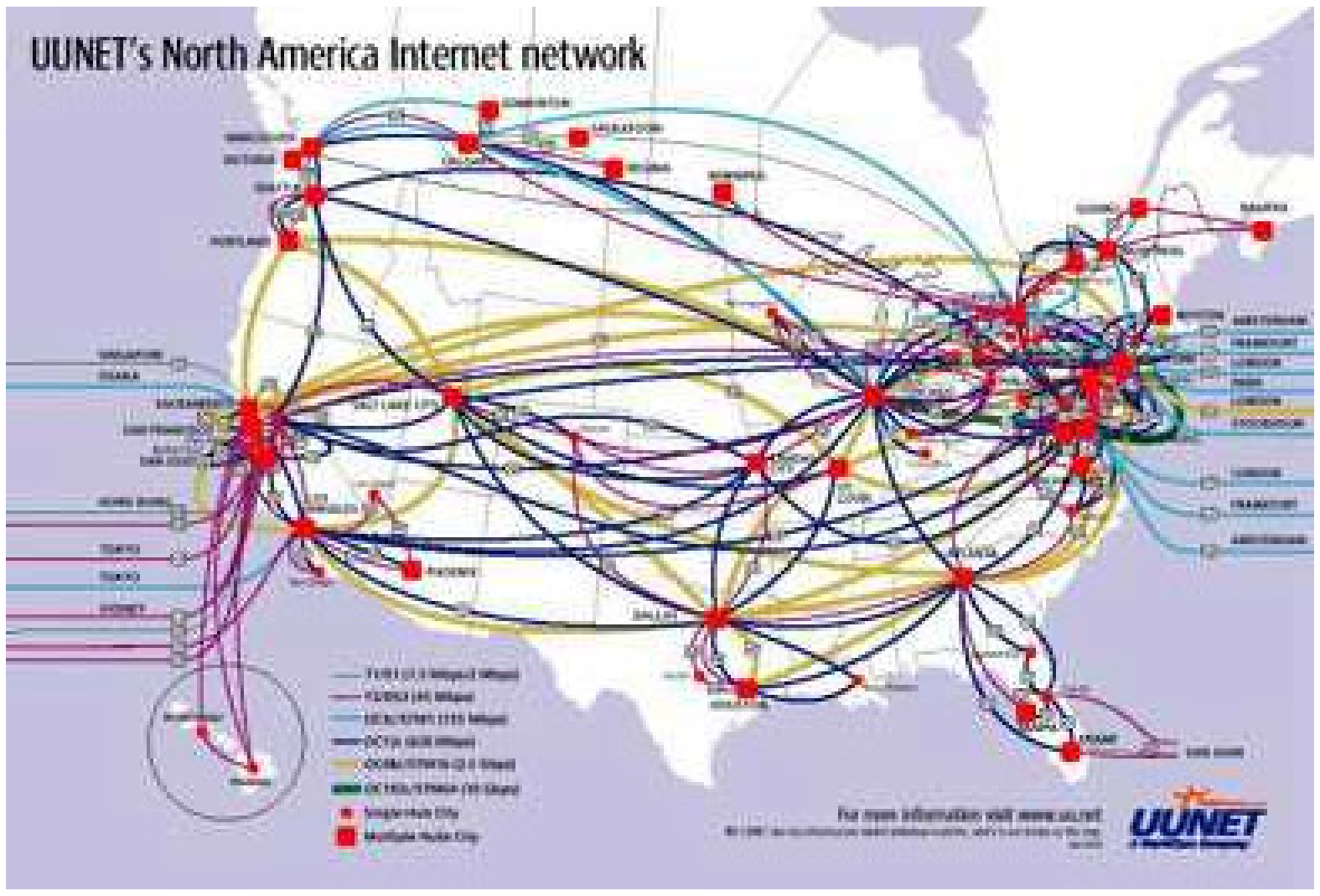

*Figure 8 System Complexity and the need for robustness http://www.cds.caltech.edu/~doyle*

When the system expresses a change to the user, the length of time taken for this is the **Response Time**. The shorter the response time the better for the user. Instantaneous response times indicate immediate reactions, e.g. pull down menus should pop up immediately as soon as they are clicked.

System complexity in almost every field is rising. With the rise of complexity comes the increasing requirement for robustness. (Illustrated in Figure 8) Intensely complex systems carry potentially increasing catastrophic risks. Failure cascade can propagate rapidly in highly interdependent systems leading to total system collapse. The requirement therefore is that robustness plays a functionally dependent role upon system complexity. Advanced systems approach the complexity of organisms which combine internal network and feedback loops to create systems which are so robust as to be apparently unperturbed by any environmental stimulus. Robustness architectures therefore need to form a foundation in the structure of any complex system.

The robustness of a system denotes its ability to withstand unexpected interruption to its normal operation or make the system less vulnerable to attack. This might be a hardware failure, a network glitch or simple operator error. Whatever the cause, an application should ideally be able to recover with minimal effort from such problems and should never lose or corrupt data or endanger health and safety in the process.

Our choice of feedback/ robustness as the fourth fundamental principle is we believe justified on three grounds

**Justification 1** Course Literature
Firstly by the importance given to it within the research literature:

Preece[31] states that, apart from "*complementary and well coordinated input devices" which must be "easy to use"...."there must also be adequate and appropriate system feedback (Norman, 1988) to guide, reassure, inform and, if necessary correct users' errors*"

Schneidermann[32] speaks of the 8 golden rules of interface design. Rule No. 5 is "*Offer error prevention and simple error handling*". This is the equivalent of recoverability within Dix's language. Schneidermann's principle 5 says, " *as much as possible design the system such that users cannot make a serious error*"

Dix et al[33] quotes robustness as one of his three central principles to support usability. He defines robustness to be composed of 1 observability, 2. recoverability 3. responsiveness and 4. task conformance. This makes robustness in Dix's language a key category. [34]

**Justification 2 Research literature**

Armen Zakaria et al in a report on Modelling Analysis of System robustness has indicated that it is essential "*to predict and anticipate possible downstream quality problems at the very early stages of system development and conduct design activities to prevent these problems from occurring.*[35]

Doyle et al[36] have observed that robustness involves "*Trade offs between a broad spectrum of environmental influences that may initiate cascading failure events. Indeed robustness can be viewed as the underlying mechanism leading to complexity*".

This importance placed upon robustness occurring at the design phase is central to our choice of this our last principle.

**Justification 3 User Survey**

Original research and a user survey has indicated quite strongly that informative feedback was the fifth most important choice out of eight selected by users. This corresponds strongly with our comparison test, which showed that a ratio of 6:4 respondents favoured "provides appropriate feedback" to "customisations". [37]

**Web site application of Feedback/Robustness**

The recipe Website has incorporated the principle in the following way: The Website contains a feedback form to provide additional comments and an email link for submission of new user recipes.

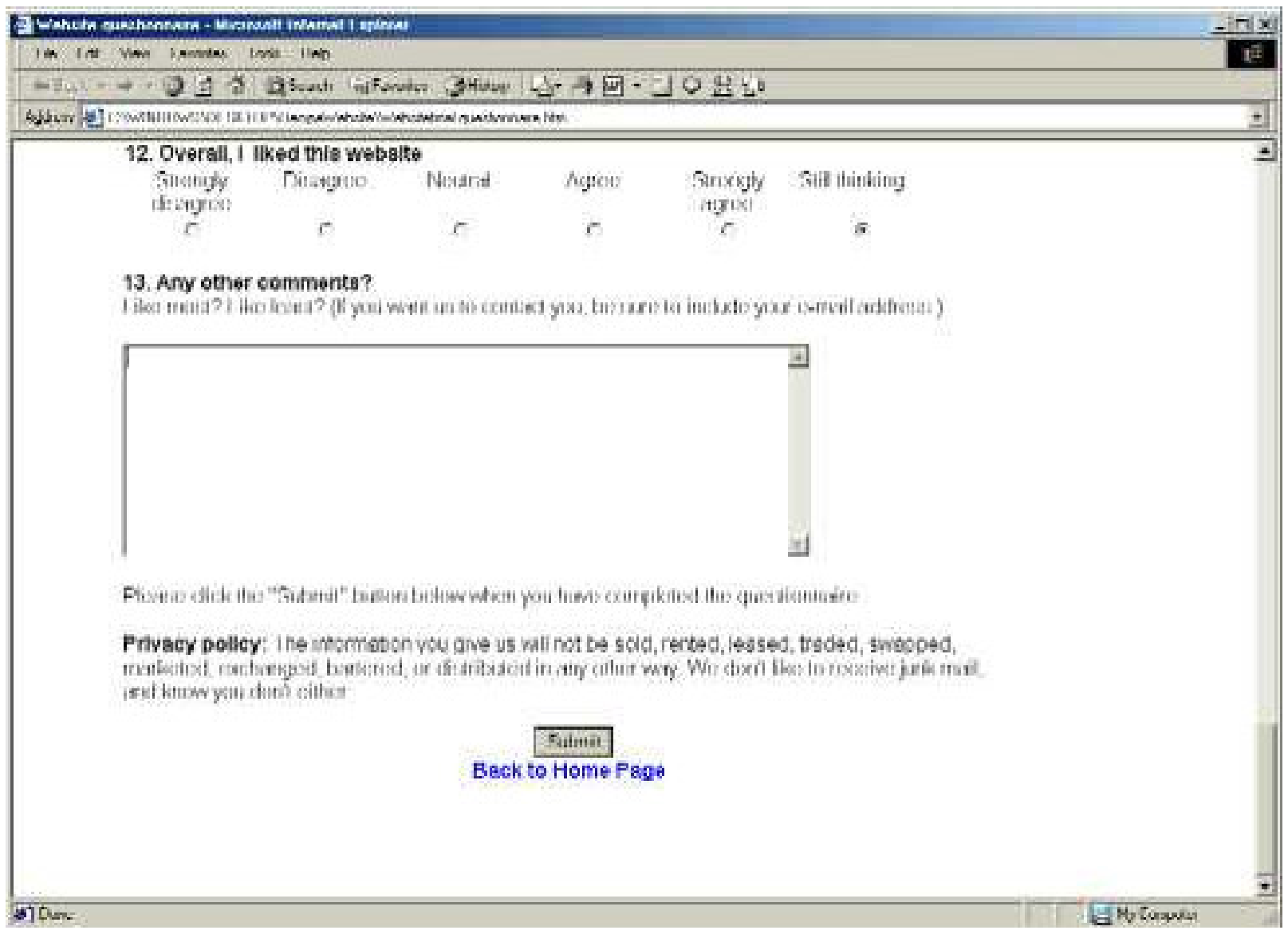

By completing the feedback form or using the email link to provide new recipes the user is providing input for the system, which is evaluated and used as output from the system when published as a new recipe for example.

The link provided at the bottom of the index page enables users to fill out an online questionnaire pertaining to the Website and the application of the four principles. By submitting questionnaires for evaluation of the Website users are providing feedback into the system, which becomes output for the Website through evaluation, which makes the human-machine-interaction cycle complete.

## Conclusion

Good design principles for human centred systems are derived from careful examination of user experience and an understanding of user cognitive psychology. These high level guidelines are based on the principle of always keeping the user at the forefront. The application of these principles has been seen to involve a process of balancing one constraint against another and refining the choices to particular contexts. This could only be done effectively when both the users' tasks and their requirements are fully understood and applied.

# APPENDIX 1

**Usability testing**: Collecting quantitative data
We conducted a usability study of the Recipe Website. Unlike many usability tests that aim to collect qualitative usability data, we wanted harder, quantitative data on which to justify some of the choices made in this essay. This meant testing 10 participants — 4 more than the number commonly used in usability testing.

### Participant sample

The work was conducted in a local educational establishment amongst a range of computer users familiar with web site navigation. We we sent out 48 requests for survey completion and recruited 10 participants, containing a balanced mix of ages (between 20 and 45 yrs), and gender.

### Websites

The prototype was a fully functional Website. In our testing, we asked the user to do three things. First, provide a rating of the most important principles (out of 8 provided) for web design. Second, compare and contrast pair-wise principles. Third, view the Website and then rate it by agreeing or disagreeing with certain statements.

### Test design

Each participant used the Recipe Website. We asked each person to perform a number of tasks with each Website in turn. After completing all the tasks participants completed a short questionnaire measuring a range of variables that our previous testing has identified as key usability measures.

### Results

Overall, the recipe Website proved easy o use, achieving the high scores in usability.

| **Criteria** | **Point score** | Ranking |
|---|---|---|
| | | |
| 1 Easy to learn | **28** | **3** |
| 2 Relevance to subject | **53** | **6** |
| 3 Familiarity of Layout | **72** | **7** |
| 4 Speed of download | **18** | **1** |
| 5 Similarity to other sites | **72** | **7** |
| 6 Consistency | **36** | **4** |
| 7 Informative Feedback | **44** | **5** |
| 8 Least amount of work to achieve result | **24** | **2** |

| **PRINCIPLE 1** | **Tick either box** | | **PRINCIPLE 2** |
|---|---|---|---|
| | | | |
| Easy to Learn | **2** | **8** | Speed of download |
| | | | |
| Full of surprises | **1** | **9** | Consistent Layout |
| | | | |
| Tested on real users | **3** | **7** | Different ways to do the same thing |
| | | | |
| Can be customized to use | **4** | **6** | Provides appropriate feedback |

| | strongly disagree | disagree | neutral | agree | strongly agree | unsure |
|---|---|---|---|---|---|---|
| 1. I would recommend this website to a friend or colleague | | 1 | 2 | 6 | 1 | |
| 2. I thought this website was hard to use | 4 | 5 | 1 | | | |
| 3. The order page was easy to use | | | 4 | 6 | | |
| 4. This website did not work very well | 3 | 7 | | | | |
| 5. The navigation of this website was intuitive and familiar | | | 2 | 2 | 6 | |
| 6. The website was easy to understand | | | 1 | 5 | 3 | 1 |
| 7. It was easy to print out chosen recipes | | | | 5 | 4 | 1 |
| 8. One click could get me anywhere I wanted | | 1 | | 3 | 6 | |
| 9. This website catered for users with little experience | | | 2 | 4 | 3 | 1 |
| 10. This website was consistent in design | | 1 | 2 | 4 | 3 | |
| 11. It didn't take long to find things on this website | | | | 5 | 5 | |
| 12. Overall, I liked this website | 1 | 1 | 4 | 3 | 1 | |

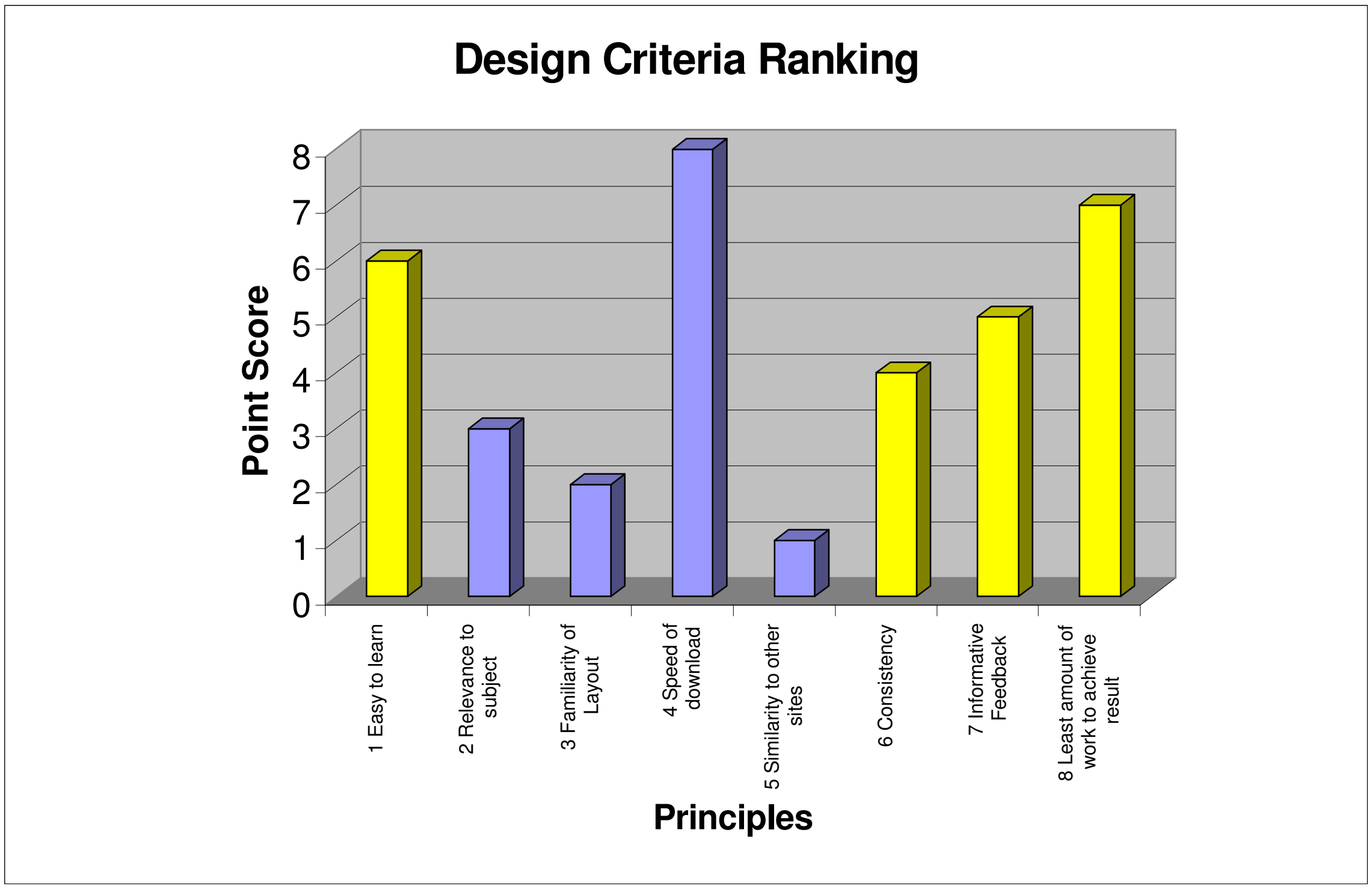

As an example, the figure above summarizes the criteria ratings provided by users. Our chosen design principles are highlighted in yellow.

# APPENDIX 2

ACMSIGCHI Curricula for Human Computer Interaction elaborates on the definition of Ergonomics as follows:

*'Ergonomics enco mpasses the wide range of Anthropometrical and physiological characteristics of people and their relationship to workspace and environmental parameters. This includes consideration of*

- *Human anthropometrics in relation to workspace design*
- *Arrangement of displays and controls, link analysis*
- *Human cognitive and sensory limits*
- *Sensory and perceptual effects of CRT and other display technologies, legibility, display design*
- *Control design*
- *Fatigue and health issues*
- *Furniture and lighting design*
- *Temperature and environmental noise issues*
- *Design for stressful or hazardous environments*
- *Design for the disabled"*[38]

Schneidermann also addresses all of the above in Chapter 1[39]